\documentclass[twocolumn,showpacs,superscriptaddress,aps,pra]{revtex4-1}
\usepackage{graphicx}
\usepackage{CJKutf8}
\usepackage{dcolumn}
\usepackage{amsmath,bm}
\usepackage{epstopdf}
\usepackage[mathlines]{lineno}
\usepackage{color}
\usepackage[colorlinks=true,linkcolor=blue,citecolor=magenta,urlcolor=cyan]{hyperref}
\begin{document}

\title{What can we learn from the experiment of electrostatic conveyor belt
for excitons?}
\author{T. T. Zhao}
\affiliation{Key Laboratory for Microstructural Material Physics of Hebei
Province, School of Science, Yanshan University, Qinhuangdao 066004, China}

\author{Rui Li}
\affiliation{Key Laboratory for Microstructural Material Physics of Hebei
Province, School of Science, Yanshan University, Qinhuangdao 066004, China}
\author{C. S. Liu}
\email{csliu@ysu.edu.cn}

\affiliation{Key Laboratory for Microstructural Material Physics of Hebei
Province, School of Science, Yanshan University, Qinhuangdao 066004, China}

\begin{abstract}
Motivated by the experiment of electrostatic conveyor belt for indirect
excitons [A. G. Winbow, \textit{et al.}, Phys. Rev. Lett. \textbf{106},
196806 (2011)], we study the exciton patterns for understanding the exciton
dynamics. By analyzing the exciton diffusion, we find that the patterns
mainly come from the photoluminescence of two kinds of excitons. The patterns near the laser spot
come from the hot excitons which can be regarded as the classical particles.
However, the patterns far from the laser spot come from the cooled excitons
or coherent excitons. Taking into account of the finite lifetime of Bosonic
excitons and of the interactions between them, we build a time-dependent
nonlinear Schr\"{o}dinger equation including the non-Hermitian dissipation
to describe the coherent exciton dynamics. The real-time and imaginary-time
evolutions are used alternately to solve the Schr\"{o}dinger equation in
order to simulate the exciton diffusion accompanied with the exciton cooling in
the moving lattices. By calculating the escape probability, we obtain the
transport distances of the coherent excitons in the conveyor which are
consistent with the experimental data. The cooling speed of excitons is
found to be important in the coherent exciton transport. Moreover, the
plateau in the average transport distance cannot be explained by the
dynamical localization-delocalization transition induced by the disorders.
\end{abstract}

\keywords{coupled quantum wells, indirect excitons, Bose-Einstein
condensation, exciton transport, dynamical localization}
\date{\today }
\maketitle

%\pacs{73.63.Hs, 78.67.De, 71.35.Lk, 73.20.Mf}

%\affiliation{Hebei Key Laboratory of Microstructural Material Physics,
%School of Science, Yanshan University, Qinhuangdao, 066004, China.}
%
%\affiliation{Hebei Key Laboratory of Microstructural Material Physics,
%School of Science, Yanshan University, Qinhuangdao, 066004, China.}

\section{Introduction}

A particle is usually called as a quantum walkers when its time evolution
follows the rules of quantum mechanics. 
Standard quantum mechanics assumes
Hermiticity of the Hamiltonian. However, non-Hermitiancity is
a universal phenomenon in many fields of physics, such as the dissipative systems with
gain and loss \cite{Ashida:2020dkc}. As a
result, the non-Hermitian quantum walker is of universal significance and
has attracted attention recently \cite{Wang_2021, PhysRevLett.128.120401,
PhysRevB.107.L140302, xiao2023observation}. The direct observation
of a quantum walker is very difficult whether it is 
in Hermitian or in non-Hermitian
systems. Taking the particle in a one-dimensional periodic potential as an
example, if the potential is  in time, the number of the particle
transport for the simplest quantum walker during the time cycle equals to
the Chern number. The Chern number characterizes the topological structure
of the model mapping from the parameter space of wave vector and time ($k,t$%
) to the momentum space ($k_x, k_y$) \cite{PhysRevB.27.6083}. The gedanken
experiment considered by Thouless demands adiabatic approximation, i.e. very slowly changing
potential. %the transport of the simplest
%quantum walkers is the Chern number according to the Thouless mechanism \cite%
%{PhysRevB.27.6083}.
%The gedanken experiment demands adiabatic approximation,
%that the potential changes very slowly.
In particular, the particles must be the Fermions occupying in a filled Bloch
bands at zero temperature.

The excitons, Bosonic quasiparticles, are the electron-hole bound pairs in
semiconductor which are expected to realize Bose-Einstein
condensation \cite{Keldysh1968}. Due to the finite lifetime, the exciton
systems have the endogenously non-Hermitianity. Both the short lifetime and
the low cooling rate hinder the realizing of the Bose-Einstein condensation.
In order to overcome the above two disadvantages, the indirect excitons,
i.e. the spatially separated electron-hole bound pairs, were generated in
coupled quantum wells \cite{Butov2002a}.
%These long-lived particles could provide a means to transport information as electrons.
As their neutral overall, they are harder to manipulate electrically. However, 
the energy of the indirect excitons can be
controlled by voltage due to the dipole moment. By
applying an alternating voltage to an electrode grid that covers the device,
a wavelike potential was created for the excitons. The excitons are
modulated by the wavelike potential which acts as the
conveyor belt across the sample. Since the high density of
excitons has the high luminous intensity,
the exciton pattern allows us to track the location of the excitons. The exciton pattern in the
moving lattices, or the conveyor belt, gives the opportunity to observe the
non-Hermitian quantum walker.

\begin{figure}[ht]
\begin{center}
\hspace*{-0.4cm} \includegraphics[width=7.0cm]{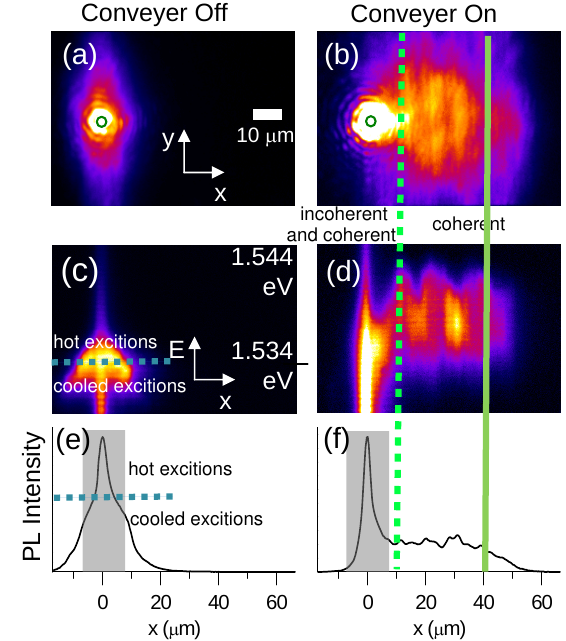}\\[0pt]
\vspace{0.8mm} %\hspace*{-1mm} %
\includegraphics[width=3.9cm]{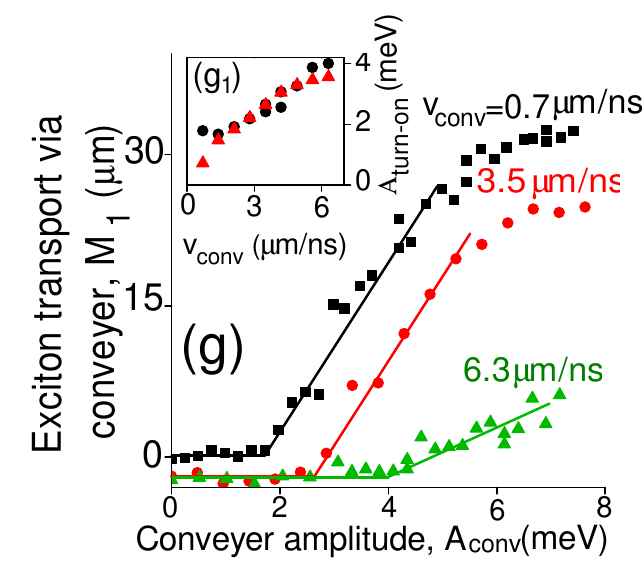} \hspace*{%
0.6mm} \includegraphics[width=3.9cm]{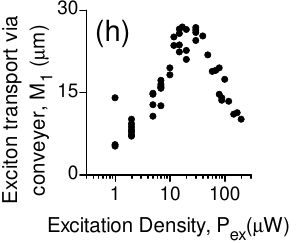}
\end{center}
\caption{The figures are taken from Ref. \protect\cite%
{PhysRevLett.106.196806}. (a-d) $x-y$ and $x-energy$ PL images, and (e, f)
PL intensity profiles $I(x)$ for conveyer off and on here $P_{\mathrm{ex}} = 20$
$\protect\mu$W, $A_{\mathrm{conv}} = 7.5$ meV, $v_{\mathrm{conv}} = 0.7$ $%
\protect\mu$m/ns. The blue dashed lines in (c) and (e) are used to
approximately separate the hot excitons near the center and the cooled
excitons far from the center when the conveyer is off. The green dashed line
across (b), (d) and (f) is used to indicate the hot (incoherent and
coherent) excitons near the center and the coherent excitons far from the
center when the conveyer is on. The green solid line across (b), (d) and (f)
indicates the cooled excitons (degenerated excitons) traveling about 40 $%
\protect\mu$m. (g) The average transport distance of indirect excitons via
conveyer $M_1$ as a function of the conveyer amplitude, $A_{\mathrm{conv}}$.
Lines are a guide to the eye. The intersection point of the line is defined
as $A_{\mathrm{turn-on}}$ where the dynamical localization-delocalization
transitions occur as is shown in (g$_1$). $P_{\mathrm{ex}} = 20$ $\protect\mu$
W. (h) The measured average transport distance of indirect excitons via
conveyer $M_1$ as a function of density. $A_{\mathrm{conv}} = 4.9$ meV, $v_{%
\mathrm{conv}} = 0.7$ $\protect\mu$m/ns. }
\label{PRL_106_196806_2011}
\end{figure}

The data of electrostatic conveyor belt for indirect excitons were reported
in Ref. \cite{PhysRevLett.106.196806}. [(a) and (b)], [(c) and (d)] and [(e)
and (f)] in Fig. \ref{PRL_106_196806_2011} show the $x$-$y$
photoluminescence (PL) images, $x$-energy PL images and $x$- PL intensity
respectively for conveyer off and on. The left column of the last row [Fig. %
\ref{PRL_106_196806_2011} (g)] shows the average transport distance of
indirect excitons via conveyer $M_1$ as a function of the conveyer
amplitude. Fig. \ref{PRL_106_196806_2011} (h) shows the distance of indirect
excitons via the conveyer $M_1$ as a function of density. The first moment
of the PL intensity $M_1 = \int x I(x) dx / \int I(x) dx$, which
characterizes the average transport distance of the indirect excitons via
conveyer. $I(x)$ is the PL intensity profile obtained by the integration of
the $x$-energy images over the emission wavelength. Major features of the
indirect exciton transport are summarized as follow. (i) There exist the
dynamical localization-delocalization transitions. (ii) Crossing the
transition point, the transport distance increases with the conveyer
amplitude and tends to saturation. (iii) The exciton transport is less
efficient for higher velocity. (iv) The efficient exciton transport via the
conveyer only occurs at intermediate densities. (v) Several bright stripes
are shown in the PL images.

The above experimental facts raise several interesting questions which need
to be clarified. For example, whether the efficiency of the exciton
transport for lower conveyer velocity is also related to the Chern number,
or not \cite{PhysRevB.27.6083}? The other important issue is whether the
several bright stripes come from the exciton coherence or not. The excitons in the coherent state or the degenerate state mean the off-diagonal long-range order exists in the system and the excitons are in a condensed state. Previously, a
nonlinear Schr\"{o}dinger equation including the attractive two-body
interaction and the repulsive three-body interaction was proposed to explain
the complex exciton patterns where the excitons are assumed to be in a
coherent state \cite{Liu2006, Xu_2012, LIU2014193, PhysRevB.80.125317}. One
may ask naturally if the nonlinear Schr\"{o}dinger equation can explain the
exciton transport. In Ref. \cite{PhysRevLett.106.196806}, the indirect
excitons were considered as classical particles. The conveyer velocity and
amplitude, as well as the exciton density, dependencies of the transport
distances are all explained by a classical diffusion equation. However, the
exciton degeneracy hidden in the PL patterns were not studied. As will be seen, the spatial and spectral patterns give the opportunity to coherence of the indirect excitons.

Since any realistic material inevitably contain a certain degree of
impurities and defects, and the interactions between particles are almost always
present, the combined effect of disorder and interactions can lead to novel
quantum states such as the Bose glass phase in disordered Bosonic systems
\cite{PhysRevLett.66.3144, PhysRevLett.98.170403}. Recently, the phase
transition between the Bose glass and the superfluid was directly observed
in the two-dimensional Bose glass in an optical quasicrystal \cite%
{yu2023observing}. The observed dynamical localization-delocalization
transitions in conveyer for excitons gives another platform to answer more
fundamental questions about the delocalization-localization transition \cite%
{High:07}.

By analysing the PL patterns, we find the exciton transport can be
simplified as the diffusion of Bosons with the finite lifetime in a moving
lattice. The main findings of this work are summarized as follow. (i) The
pattern can be divided approximately into the incoherent part and coherent
part which can be described by the diffusion differential equation used in
Ref. \cite{PhysRevLett.106.196806} and the nonlinear Schr\"{o}dinger
equation proposed by us respectively. (ii) The bright stripes are the
interplay between the coherent excitons and the moving lattices. (iii) The
cooling rate is the key factor in the exciton transport. (iv) The lifetime
and cooling rate of the coherent excitons are estimated. (v) The sample is
very clean and free of impurity. We discuss the data and build a
time-dependent nonlinear Schr\"{o}dinger equation in Sec.~\ref{Model
Hamiltonian} to describe the coherent excitons. We also show how to obtain the
PL intensity profiles $I(x)$ by calculating the escape rate. In Sec.~\ref%
{Numerical analysis}, numerical calculations and detailed discussions on the
exciton distribution in moving lattices are given. Sec.~\ref{Summary} is
devoted to a brief summary.

\section{Model Hamiltonian}

\label{Model Hamiltonian}

The clues of understanding the PL patterns in the conveyor belt come from
the various exciton patterns found by the previous experiments, in
particular, the two puzzling exciton rings, inner and external, and periodic
bright spots in the external ring \cite{Butov2002a, Butov2002b, Snoke2002a}.
A charge-separated transport mechanism was proposed and gave a satisfactory
explanation both of the formation of the two exciton rings and of the dark
region between the inner and the external ring \cite{butov:117404,
rapaport:117405}. This mechanism was further confirmed by PL images of
single quantum well \cite{rapaport:117405}.

As pointed out in Refs.~\cite{butov:117404, rapaport:117405}, when electrons
and holes are first excited by high-power laser, they are actually
charge-separated and have a small recombination rate. No true excitons are
formed at this stage. They can travel a long distance from the laser spot
before recombination. After a long-distance traveling, the hot electrons and
holes collide with the semiconductor lattices and are cooled down. The
cooled electrons and holes forms the excitons in the annular region near the the laser spot. The recombination of this part of excitons generates a PL ring.
As the drift speed of electrons (with smaller effective
mass) is larger than that of holes (with larger effective mass), there are always some
electrons and holes escaping from this recombination. Also due to the
neutrality of the coupled quantum wells, the negative charges will slow down
and begin to accumulate at the sites far away from the PL ring. The cold
electrons and holes then eventually meet to form the excitons. This part of the excitons recombine and also form the exciton ring
at the boundary of the opposite charges. The second ring is far away the laser spot comparing with the former. The two rings were called as inner ring and external ring respectively.

As the excitons forming in the external ring are from the cooled electrons
and holes, they have a low kinetic energy and temperature. While whether
these excitons are condensed is in debate \cite{butov:117404,
PhysRevLett.93.266801, PhysRevLett.97.016803, doi:10.1021/nl300983n,
doi:10.1021/nl302504h, PhysRevLett.94.176404, PhysRevB.74.085303,
PhysRevB.76.115303}, we reasonably assumed that the excitions were in the
highly degenerated states, and proposed a self-trapped interaction model
involving an attractive two-body interaction and a repulsive three-body
interaction \cite{Liu2006, Xu_2012, PhysRevB.80.125317}. The mechanism gave
a good account to the periodic bright spots in the external ring. In
addition, it also explained well the abnormal exciton distribution in an
impurity potential, in which the PL pattern became much more compact, and exhibited an annular shape with a
darker central region \cite{LaiCW2004}. Moreover, the model also captured
some experimental details. For instance, the dip can turn into a tip at the
center of the annular cloud when the sample was excited by a higher power
lasers.

Inspired by both the above experimental data and theories, we are ready to
investigate the exciton PL patterns in the moving lattices. As we know, the PL
intensity is approximately proportional to the exciton number and the PL energy is
approximately proportional to the exciton energy which includes the kinetic
, the bound and the interaction energies. However, the bound
energy remains unchanged due to the special structure of the
spatially separated electrons and holes. The interaction energy between
excitons is also negligible in the low-density case. As a result, the change
of PL energy is only related to the kinetic energy. The PL energies shown
in Fig. \ref{PRL_106_196806_2011} (c) (1.534 eV marked by blue dashed line)
and PL intensities in Fig. \ref{PRL_106_196806_2011} (e) far from the center
are lower than those near the center. Therefore, it indicates that the excitons near
the center have larger kinetic energy and higher particle density. We argue
that the spatial distribution of excitons here is also consistent with that
of the exciton rings reported in Ref. \cite{Butov2002a, Butov2002b}. The hot
excitons are formed at the center [dark region -10 $\mu $m $<$ x $<$ 10 $\mu$%
m in Fig. \ref{PRL_106_196806_2011} (e)] and the cooled excitons are located far
from the center. So there exist two kinds of excitons, one is the hot
excitons near the center and can be regarded as classical particles, the other
is the cooled excitons far away the center cannot be regarded as classical
particles.

%whether they are in coherent state or not,
When the conveyer is turned on [see Fig. \ref{PRL_106_196806_2011}
(b)], the moving potential modifies both the exciton potential energy and the PL
energy [see Fig. \ref{PRL_106_196806_2011} (d)]. While the moving
lattices drag the hot excitons, they are cooled down further by inelastic
collision to the semiconductor phonons. In comparison with the high
temperature case near the center, the excitons far from the center have a long de Broglie
wavelength in the low temperature region. As we know, a
quantum phase transition occurs when the Bosons condense from the
non-degenerate state to degenerate state. Whether the finite life-time
excitons can experience a Bose-Einstein condensation transition is an interesting issue.
Neglecting the above complexity, we roughly divide the exciton distribution
into two parts by the green dashed line shown in Fig. \ref%
{PRL_106_196806_2011} (b), (d) and (f). On the left side of the green dashed
line, part of the excitons are in coherent state and part of excitons are in
incoherent state. On the right side of the green dashed line, however, we
consider all the excitons are in coherent states or degenerate states.
According to the above analysis, the coherent excitons may come from two
parts. One part directly comes from the hot exciton near the center by the cooling.
The other comes directly from the pairing of the cooled electrons and holes.

It is useful to estimate the exciton parameters according to the above
analysis. As is indicated by the green dashed line in Fig. \ref%
{PRL_106_196806_2011}, after a 10 $\mu$m traveling, the hot excitons can
become highly degenerated excitons. As is also marked by a blue solid line,
after another 30 $\mu$m traveling, the coherent excitons recombine.
If the excitons follow the conveyor belt with a low velocity $v_{\text{conv}}=0.7
$ $\mu \text{m/ns}$, it takes about 14 ns for the hot excitons to the
highly degenerated excitons. So the cooling time of the excitons from classical particles to coherent particles is about 14
ns. It takes another 40 ns
before the degenerated excitons recombine. So the lifetime of the coherent
excitons is estimated to be about 40 ns. This is the first time to obtain
the lifetime of indirect exciton experimentally.

Based on the above analysis, it is reasonable to model the highly
degenerate exciton gas by a time-dependent nonlinear Schr\"{o}dinger
equation \cite{Liu2006, PhysRevB.80.125317}
\begin{equation}
i\frac{d\left\vert \Psi \left( t\right) \right\rangle }{dt}=H\left\vert \Psi
\left( t\right) \right\rangle ,
\label{time-dependent nonlinear Schrodinger equation}
\end{equation}%
where $\left\vert \Psi \right\rangle =\left\vert \psi _{1},\psi _{2},\cdots
,\psi _{L}\right\rangle $ and $L$ is the lattice length. The tight-binding
Hamiltonian reads%
\begin{eqnarray}
H &=&\sum_{j}\left\{ \tilde{t}\left( \left\vert \psi _{j}\right\rangle
\left\langle \psi _{j+1}\right\vert +\text{h.c.}\right) \right.
\label{The tight-binding
Hamiltonian} \\
&&+\left. \left[ V_{\text{conv},{j}}\left( t\right) +F\left( n_{j}\right)
-i\gamma \right] \left\vert \psi _{j}\right\rangle \left\langle \psi
_{j}\right\vert \right\} ,  \notag
\end{eqnarray}%
where
\begin{eqnarray}
V_{\text{conv},j}\left( t\right) &=&A_{\text{conv}}\cos \left[ 2\pi \left(
j-v_{\text{conv}}t\right) \right/ \lambda ]  \notag \\
&=&A_{\text{conv}}\cos \left( 2\pi \kappa j-\omega t\right)
\label{the moving lattices}
\end{eqnarray}%
denotes the moving external potential created by a set of ac gate voltages. When $%
v_{\text{conv}}=0$ and $\omega =0$, the moving lattice reduces to the static
lattice. The parameter $\gamma =1/\Gamma >0$ is the loss rate and $\Gamma $ is the
exciton average lifetime. $\tilde{t}$ is the hopping amplitude and is set to be
1 in the following discussions.
$n_{j}=\left\vert \psi _{j}\right\vert ^{2}$ is the local probability
density. $F\left( n_{j}\right) $ is the effective interaction between the
indirect excitons. 

In general, the excitons are considered as the weak
repulsive interactions $F\left( n_{j}\right) =gn_{j}$. Note that $F\left( n_{j}\right)
=-g_{1}n_{j}+g_{2}n_{j}^{2}$ was used to explain various exciton
patterns \cite{Liu2006, Xu_2012, LIU2014193, PhysRevB.80.125317}. The above
phenomenological interaction may come from the dipolar interaction and the
exchange interaction. Since the effective parameter $g_{1}\propto N$ and $%
g_{2}\propto N^{2}$ ($N$ is the particle number), we have the relationship $%
g_{2}\propto g_{1}^{2}$. We take $g_{2}=\epsilon g_{1}^{2}$ here $\epsilon $
is the parameter describing the complex interactions. As will be seen in Sec.~\ref%
{Numerical analysis} B,
the effects of these
two kinds of interactions are identical except for the cases of the very low
and very high particle density. Eq.~(\ref{time-dependent nonlinear Schrodinger
equation}) is the modified Gross-Pitaevskii equation when $\gamma =0$.
Unlike the Gross-Pitaevskii equation which is usually used to determine the
ground states of a low-temperature Bosonic gas with a short- or zero-range
two-body interaction, Eq.~(\ref{time-dependent nonlinear Schrodinger
equation}) contains the effect of the finite particle lifetime.

As the norm of the wavefunction decreases with time
\begin{eqnarray*}
\frac{d}{dt}I(t) &=&\frac{d}{dt}\langle \Psi (t)|\Psi (t)\rangle
=i\left\langle \Psi (t)\right\vert (H^{\dag }-H)\left\vert \Psi
(t)\right\rangle \\
&=&-2\gamma \langle \Psi (t)|\Psi (t)\rangle =-2\gamma I(t),
\end{eqnarray*}%
the time-dependent wavefunction norm can be written as
\begin{eqnarray}
I(t)=\langle \Psi (t)|\Psi (t)\rangle =\sum_{j}|\psi _{j}(t)|^{2}=\exp
\left( -2\gamma t\right),  \label{the time-dependent wavefunction norm}
\end{eqnarray}
when the initial-state is normalized $\langle \Psi (0)|\Psi (0)\rangle =1$. $\gamma$ is the lost rate which is inversely proportional to the life-time of the excitons. We define the escaping probability at location $j$ of the walker as
\begin{equation}
I_{j}=2\gamma \int_{0}^{\infty }dt|\psi _{j}(t)|^{2},
\label{the probability of the walker escapes from location j}
\end{equation}%
which satisfies $\sum_{j}I_{j}=\sum_{j}2\gamma \int_{0}^{\infty
}dt|\psi _{j}(t)|^{2}=\int_{0}^{\infty }dt I(t)=1$. $I_{j}$ is the summation of
luminescence intensity of all the exciton states and is the PL intensity profile.

The cooling of the excitons is another important physical factor in the exciton
diffusion. With the increase of the time, the high-energy excitons relax to low-energy
excitons and condense to their degenerate states. The condensed excitons in the confined well occupied 
the highly degenerate states can be further cooled before recombination. The cooling process can be well simulated by the imaginary time evolution. 
As we know, for
a stationary state or eigenstate, if the time $t$ is artificially replaced
by $-it$, the wavefunction will exponential decay in its time evolution. According to the superposition principle, a state can be expanded as 
the linear superposition of all the eigenstates. In the imaginary time
evaluation of the state, the components with the higher eigen energies have
the faster decay rates which is equivalent to the relaxation of high energy components to the low energy
components \cite{10.1063/1.4821126}. Although the
energy levels are not well defined in the time-dependent non-Hermitian
system, the energy levels have previously been observed in previous experiments \cite%
{Butov2002a, Butov2002b, Snoke2002a, LaiCW2004, high-2008} and investigated
theoretically \cite{1742-6596-210-1-012050, Liu2006, Xu_2012, LIU2014193,
PhysRevB.80.125317}. We use the imaginary time-evolution of the Schr\"{o}dinger
equation in Eq.~(\ref{time-dependent nonlinear Schrodinger equation}) to
describe the cooling process.

%To simplify the discussions, the excitons are assumed to
%be in the Boltzmann distribution $\sim e^{-\beta E_{n}}$ in which $E_{n}$ is
%the energy level and $\beta=1/{k_{B}T}$. $k_{B}$ and $T$ are the Boltzmann
%constant and the temperature respectively.

In the following calculations, we solve the time-dependent Schr\"{o}dinger
equation (\ref{time-dependent nonlinear Schrodinger equation}) in a real time interval $\Delta
t$ with the initial state $\Psi (0)$ to obtain the time evolution state $\Psi (\Delta t)$. Then the state $%
\Psi (\Delta t)$ is used as the initial state, we solve the time-dependent Schr\"{o}dinger
equation (\ref{time-dependent nonlinear Schrodinger equation}) in a imaginary time interval $i\Delta
t^{\prime }$ to obtain the time evolution state  $\Psi (\Delta t+i\Delta t^{\prime })$. At last, the
wavefunction $\Psi (\Delta t+i\Delta t^{\prime })$ is normalized as \cite%
{10.1063/1.4821126}
\begin{eqnarray}
I(\Delta t+i\Delta t^{\prime }) = |\Psi (\Delta t+i\Delta t^{\prime })|^2
%&=&\langle \Psi (\Delta t+i\Delta t^{\prime })|\Psi (\Delta t+i\Delta
%t^{\prime })\rangle  \notag \\
=\exp \left( -2\gamma \Delta t \right) .
\end{eqnarray}%
Repeating the above process $\mathcal{N}$ times until $I\left( \mathcal{N}%
\left[ \Delta t+i\Delta t^{\prime }\right] \right) \sim 0$, we get a serial of $%
\Psi \left( n\left[ \Delta t+i\Delta t^{\prime }\right] \right) $ to calculate $I_{j}=2\gamma \sum_{n=0}^{\mathcal{N}}\left\vert \Psi \left( n\left[
\Delta t+i\Delta t^{\prime }\right] \right) \right\vert ^{2}$. We define $%
\tau =\Delta t^{\prime }/\Delta t$. The larger $\tau $ indicates the longer
cooling time. So $\tau $ is proportional to cooling speed.

It should be emphasized that the method above is effective since the
normalization condition of the escape rate $I_j$ in Eq. (\ref{the
probability of the walker escapes from location j}) guarantees the
convergence of the calculations. Otherwise, without considering the
time-dependent normalized condition, the propagation of a wave-pocket still
meets the boundary and reflects under the open boundary condition. Even though the
absorbing boundary condition is adopted, the boundary also have a great effect on the evolution of the wavefunction. In such a case, the interference fringes are formed
by interference between incident and reflected waves and no
convergent solution can be obtained.

\section{Numerical analysis}

\label{Numerical analysis}

\subsection{A wave pocket diffusion: without particle interactions}

\begin{figure}[htbp]
\hspace*{-0.32cm} \includegraphics[width=8cm]{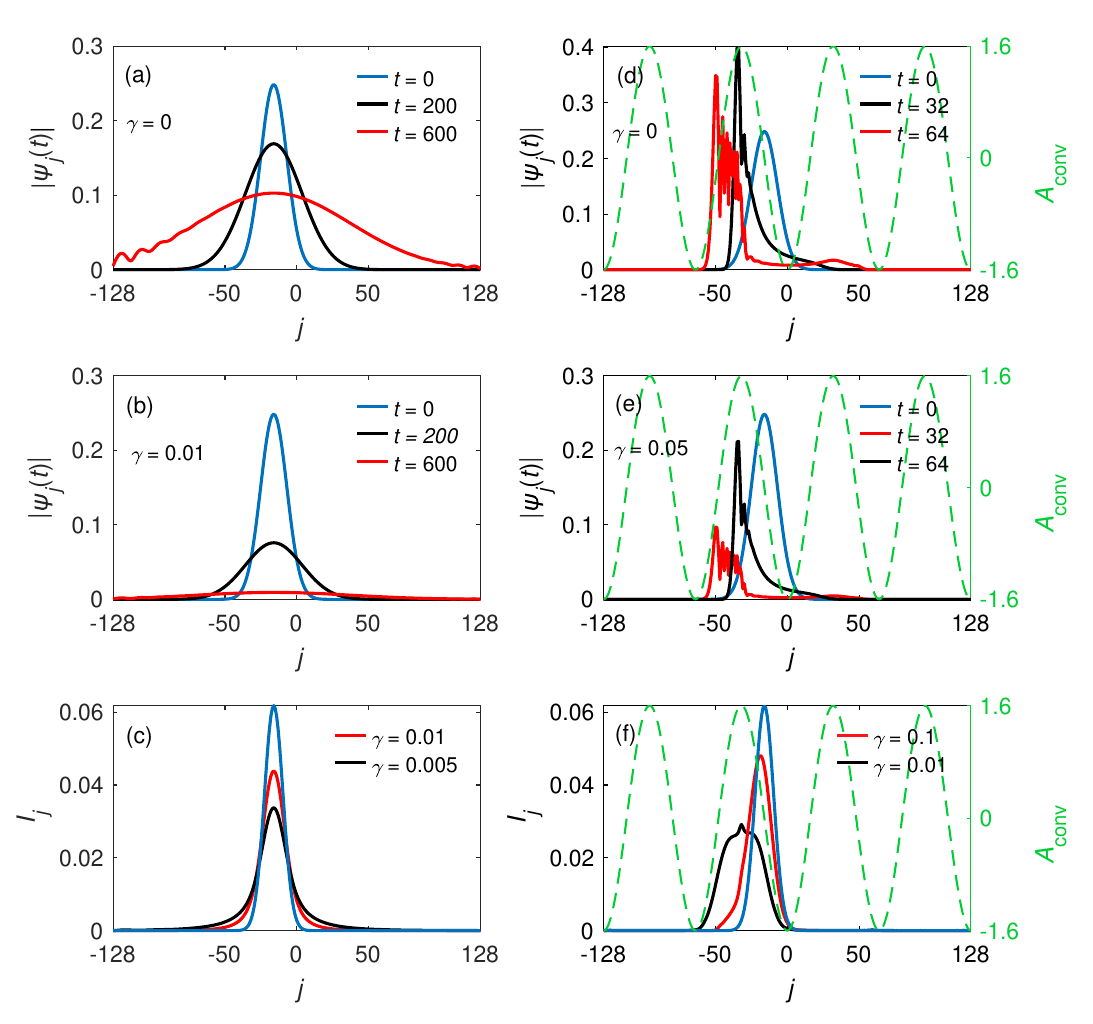}
\caption{The wave pocket diffusion without a periodic potential (left
column) and with the static periodic potential (right column). The free
diffusion of a wave pocket at different time without the dissipation $%
\protect\gamma =0$ (a) and with the dissipation $\protect\gamma =0.01$ (b).
The diffusion of a wave pocket modulated by the static periodic potential
(dashed line) at different time without the dissipation $\protect\gamma=0 $
(d) and with the dissipation $\protect\gamma=0.05$ (e). $I_j$ distribution
for the different dissipation $\protect\gamma$ for the free diffusion (c)
and for the diffusion modulated by the static periodic potential $V_{\text{%
conv},j}$ (f). The parameters of the periodic potential in Eq. (\protect\ref%
{the moving lattices}) are set as $A_{\text{conv}} = 1.6$, $\protect\omega%
=0$, and $\protect\lambda = 64$.}
\label{wave_pocket_diffusion}
\end{figure}

According to the above analysis, there are various sources for the coherent excitons. As a result, the initial distribution of the degenerate excitons can be
considered as a wave pocket with multi-peaks. In order to understand the 
transport of the wave pocket, let us first study the simplest case of the diffusion of a
wave pocket with one-peak. We solve the time-dependent nonlinear Schr\"{o}%
dinger equation (\ref{time-dependent nonlinear Schrodinger equation}) in
real-time only under the open boundary condition ($\psi _{\pm L}=0$) for an
initial Gauss wave pocket $\psi _{j}(0)=\sqrt{(2\kappa /\pi )}\exp \left[
-\kappa (j-j_{0})^{2}\right] $. The results with the parameters $j_{0}=-16$,
$\kappa =0.006$ and $L=128$ are shown in Fig. \ref{wave_pocket_diffusion}.
In the following discussions, the blue lines in all the figures indicate the shape of the
initial wave pocket.

The physics of the wave pocket diffusion is very simple \cite{zeng2007,
Sakurai1994}. For the Gauss wave pocket, its wave functions both in real space
and momentum space are all of the Gauss type. According to the uncertainty relation
$\Delta x\Delta p\approx \hbar $, with the increase of time, the larger $\Delta x$
leads to the smaller $\Delta p$, here $\Delta x$ and $\Delta p$ are the full width
at half maximum of the Gauss wave pocket. If we approximately take $v
\approx \Delta v$, the smaller $\Delta p$ leads to the lower the propagation
speed. As a result, although the diffusion becomes slower and slower as the time
increases, the wave finally arrives at the boundary and is reflected.
Interference fringes are generated by the interference between reflected and
incident waves. The above statements are true whether it is without the
periodic potential in Fig. \ref{wave_pocket_diffusion} (a) or with the
static periodic potential in Fig. \ref{wave_pocket_diffusion} (d). The
difference is that more interference fringes appear in Fig. \ref%
{wave_pocket_diffusion} (d) due to the interplay between the wave and the
periodic potential. When we take the dissipation of the particles into account, the
wave pocket decays in Fig. \ref{wave_pocket_diffusion} (b) and (e) with the
time goes on. The escape probability $I_{j}$ in Eq. (\ref{the probability of
the walker escapes from location j}) is calculated and shown in Fig. \ref%
{wave_pocket_diffusion} (c) and (f). As expected, the particle dissipation
suppresses the wave pocket diffusion. From Fig. \ref{wave_pocket_diffusion}
(f), $I_{j}$ can be deviated from the initial position due to the modulation
of the external potential.

\begin{figure}[htbp]
\includegraphics[width=8cm]{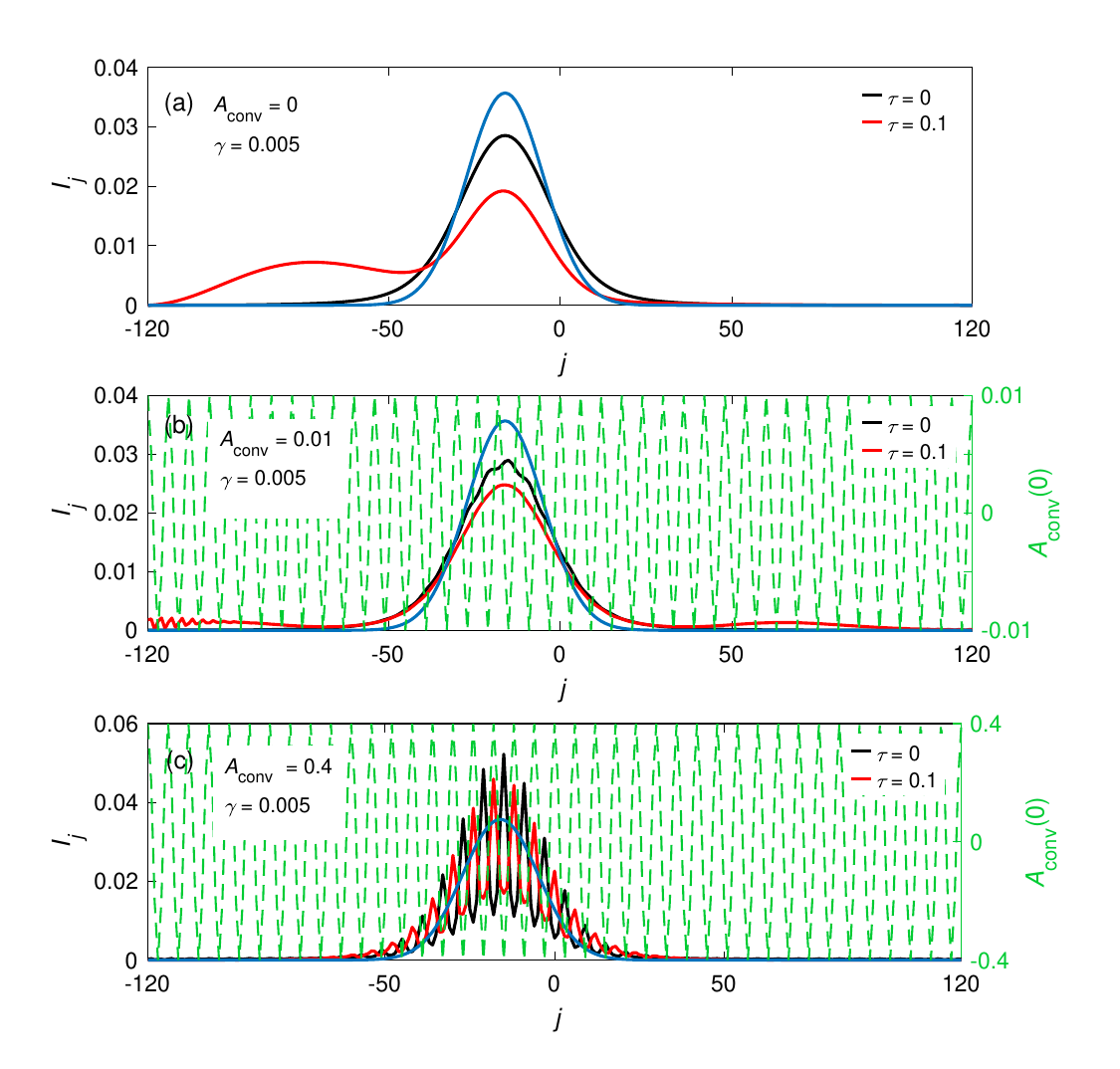}
\caption{The wave pocket diffusions without the cooling [black lines] and  with
the cooling [red lines] modulated by the static periodic potentials with the
different intensities (a) $A_{\text{conv}} = 0$, (b) $A_{\text{conv}} = 0.01$
and (c) $A_{\text{conv}} = 0.4$. }
\label{wave_pocket_diffusion_cooling}
\end{figure}

We next investigate the cooling effect on the wave pocket diffusion. 
The time-dependent nonlinear Schr\"{o}dinger equation (\ref{time-dependent
nonlinear Schrodinger equation}) is solved with the real-time and imaginary-time evolutions alternatively to obtain the escape rate $I_{j}$ of the coherent excitons. The result of $%
I_{j}$ is shown in Fig. \ref{wave_pocket_diffusion_cooling}. It can be seen that
the cooling helps the wave pocket diffusion. We can understand the
underlying physical picture as follow \cite{zeng2007, Sakurai1994}. A
wave pocket in free space can be considered as the superposition of the
plane waves with the different wave vectors. When the condensed Bosons are
considered to be in a state of Gauss wave pocket in momentum space, their distribution in the
Cartesian coordinate space is still Gauss type. With the time increases, the
wave pocket will spread in coordinate space and will contract in momentum
space. However, their waves are still Gauss types. With the temperature
decreases, the low-energy probability of the Gauss wave pocket increases
accordingly since the wave pocket in momentum space is modulated by the
factor $\exp (-\beta E_{k})$ ($E_{k}=k^{2}/2$). The narrower the wave pocket in
momentum space, the wider wave pocket in coordinate space \cite%
{zeng2007, Sakurai1994}. As a result, the quantum mechanical effect is in
opposition to the classical effect where the spreading of classical particle
becomes slow with the decrease of the temperature.

%When the wave pocket is modulated by static periodic potential, the
%particles spread must cross the potential barrier. With increasing the
%periodic potential $A_{\text{conv}}$, the low tunneling probability hinders
%the wave pocket spread.
By comparing the black lines and the red lines in Fig. \ref%
{wave_pocket_diffusion_cooling} (b), it is found that the cooling also
helps the wave pocket spread in a static periodic potential. The physical
picture can also be understood in a similar way. In the static periodic
potential, the eigenfunction of a particle is the Bloch wave which can be
expanded in terms of the Wannier functions. Correspondingly, the energy bands
come from the splitting of the degenerate energy levels. A wave pocket in
periodic potential can be considered as a superposition of the Bloch waves
with the different wave vectors. With time increasing and particle cooling, the
particle probabilities in low-energy bands increase accordingly. The
narrower the wave pocket in momentum space, the wider the wave pocket in coordinate
space. As a result, the cooling enhances the diffusion of the wave pocket.

It is interesting to see in Fig. \ref{wave_pocket_diffusion_cooling} (c)
that the peaks of the $I_{j}$ for $\tau >0$ are located at the crests
of the periodic potential, which is in contrast to the case of $\tau =0$
where the peaks are located at the troughs. The reason can be understood as follows. According to the above analysis, with the lowering of the temperature, the Bosons
tend to condense at $k=0 $, which results in the increasing of $\Delta x$.
In the large $A_{\text{conv}}$ case, the Wannier functions are the atomic
orbital wavefunctions in a unit cell and the particles are confined in a
single unit cell. The stronger the confinement, the larger the $\Delta x$. The
particles tend to move to the boundary of the unit cell to obtain the large $%
\Delta x$ which causes the dip of particle distribution at the center of the
unit cell. This quantum behavior can be used to detect the
exciton degeneracy.

\begin{figure}[htbp]
\includegraphics[width=8cm]{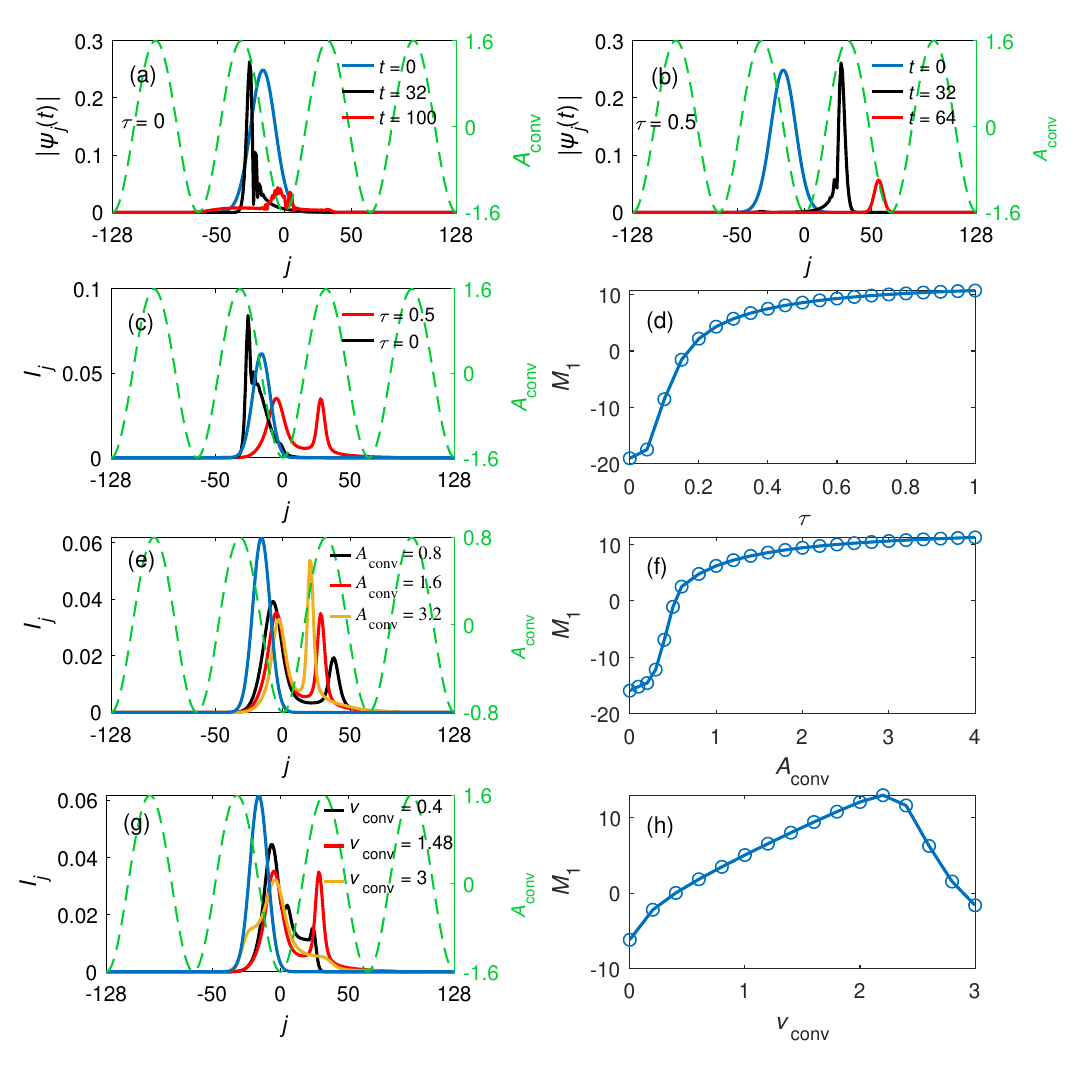}
\caption{The wave pocket diffusions in a moving lattice without considering the cooling
(a) and considering the cooling (b). $I_j$ of the two cases of (a) and (b) are
given in (c). (e) and (g) show $I_j$ with the different moving lattice
intensity $A_{\text{conv}}$ and moving speed $v_{\text{conv}}$. The cooling
speed $\protect\tau$, the lattice intensity $A_{\text{conv}}$ and the
lattice velocity $v_{\text{conv}}$ dependencies of the transport distant $M_1
$ are shown in (d), (f) and (h) respectively. The parameters used in all figures are: $A_{\text{conv}}$ = 1.6, $v_{\text{conv}}$ = 1.48, $\protect\gamma=0.05$.}
\label{wave_pocket_spreading_in_moving_lattices}
\end{figure}

The wave pocket diffusion in a moving lattice is shown in Fig. \ref%
{wave_pocket_spreading_in_moving_lattices}. The moving
lattice modifies the shape of the Gauss wave pocket. With the increase of the time, the wave pocket follows the moving lattice while its height decreases
obviously due to the finite lifetime of the exciton [Fig. \ref%
{wave_pocket_spreading_in_moving_lattices} (a) and (b)]. The exciton cooling
($\tau =0.5$) in Fig. \ref{wave_pocket_spreading_in_moving_lattices} (b)
dramatically increases the transport distance in comparison with the case
without the cooling ($\tau =0$) in Fig. \ref%
{wave_pocket_spreading_in_moving_lattices} (a). The escaping probability $I_{j}$ corresponding to the
case in Fig. \ref{wave_pocket_spreading_in_moving_lattices} (a) and (b) is
presented in Fig. \ref{wave_pocket_spreading_in_moving_lattices} (c). The
average transport distance $M_{1}$ increases with the cooling parameter $%
\tau $ and then tends to saturation as is shown in Fig. \ref%
{wave_pocket_spreading_in_moving_lattices} (d). We can conclude that the
cooling is the key factor to the exciton transport via the conveyer.

The reason can be understood from the case of the low velocity and  the high $A_{\text{conv}}$ of the
moving periodic potential. In such case, the Bloch theory can be assumed to be correct
approximately and the Wannier functions can
be approximated as the atom-orbital functions (particle wave functions in
unit cell). The overlap of the neighbouring low orbital functions is less
than that of the high orbital functions. The little overlap of the
neighbouring low orbital functions leads to the little tunneling between
unit cells. With the cooling, the excitons tend to occupy the low
energy bands. The occupation probability of the low orbital state increases with
the cooling parameters $\tau $. With increase of the cooling further, most part of
the particles are in the low orbital state. As a result, the particles follow the
moving lattice. So the change of the average transport distance $M_{1}$
obvious increase with $\tau $ as is shown in Fig. \ref%
{wave_pocket_spreading_in_moving_lattices} (d).

The tunneling between unit cells decreases with the increase of the
lattice height $A_{\text{conv}}$. In the high $A_{\text{conv}}$ case, the
tunneling between unit cells is inconspicuous. So the transport distance
increases with the $A_{\text{conv}}$ and then tends to saturation as is shown
in Fig. \ref{wave_pocket_spreading_in_moving_lattices} (e). In the low
lattice speed case, the average transport distance $M_{1}$ increases with
the lattice speed as is shown in Fig. \ref%
{wave_pocket_spreading_in_moving_lattices} (h), which indicates the exciton
transport follows the moving lattices. In the high lattice speed case, the
energy band theory breaks down. It is definitely that the tunneling between
the unit cells decreases with the increase of the lattice speed which is
adverse to the particles transport. We therefore argue that the particles
follow the moving lattices only in the low velocity and high lattice
amplitude cases.

\subsection{A wave pocket diffusion: with particle interactions}

\label{A wave pocket transport with the particle interactions}

\begin{figure}[htbp]
\includegraphics[width=8cm]{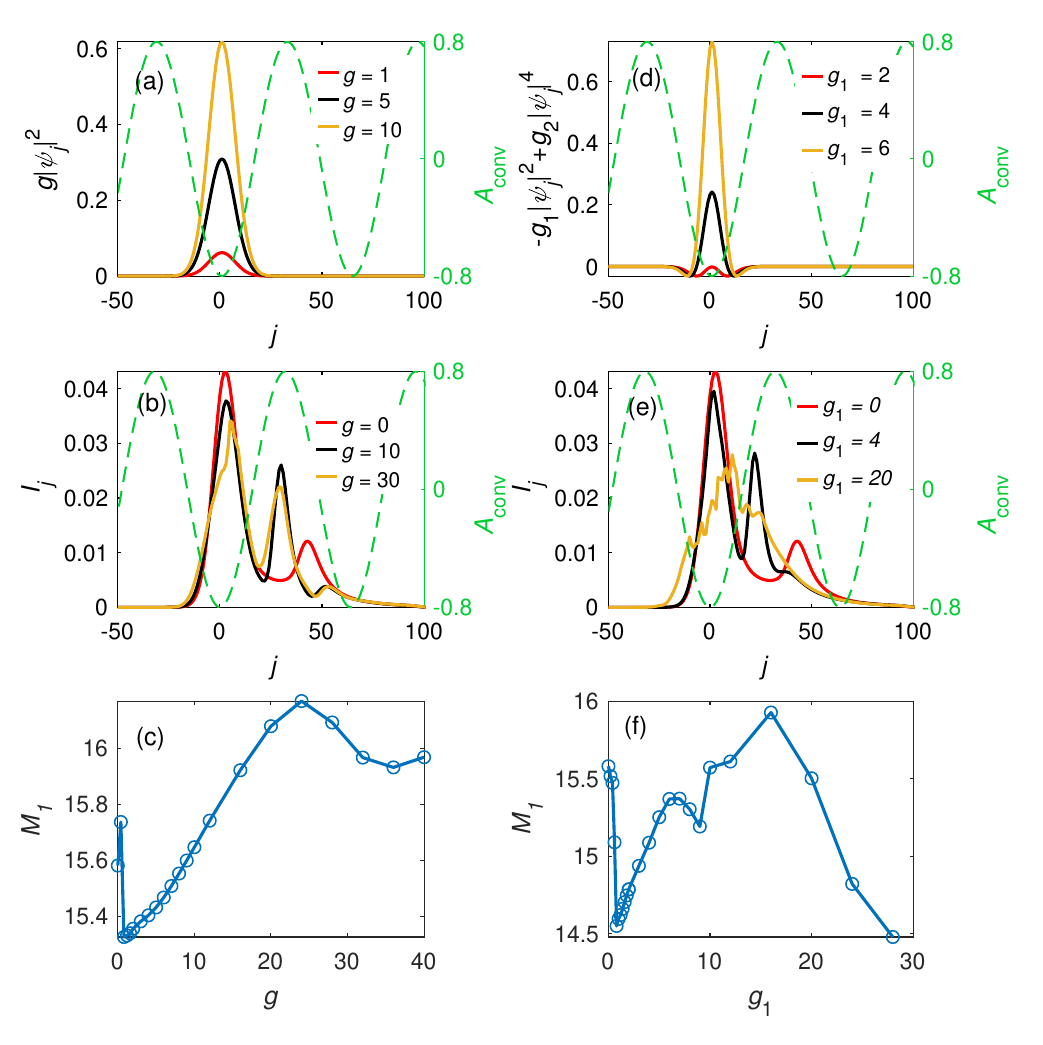}
\caption{The wave pocket diffusions of the weak repulsive particles. The
following parameters are used in the numerical analysis: $A_{\text{conv}} =
0.8$, $v_{\text{conv}} = 2.08$, $\protect\tau = 0.5$ and $\protect\gamma =
0.05$. For the pure repulsive case, the interaction parameter $g$-dependent effective potential $g|\protect\psi_j|^2$, the distribution patterns $%
I_j$ with different $g$ and transport distance $M_1$ as a function of $g$ are
given in (a), (b) and (c) respectively. For the complex interaction $-g_1 |%
\protect\psi_j|^2 + g_2 |\protect\psi_j|^4$ case, we set $g_{2}=\protect%
\epsilon g_{1}^{2}$ and $\protect\epsilon = 8$. The interaction parameter $%
g_1$-dependent effective potential, the distribution patterns $I_j$
with the different $g_1$ and transport distance $M_1$ as a function of $g_1$
are given in (d), (e) and (f) respectively.}
\label{wave_pocket_transport_repulsion}
\end{figure}

In general, the excitons are considered to have the weak repulsive interactions.
We study the exciton transport with this kind of interactions in Figs. \ref%
{wave_pocket_transport_repulsion} (a), (b) and (c). The interaction $g|\Psi
_{j}|^{2}$ acts as the effective potential as is shown in Fig. \ref%
{wave_pocket_transport_repulsion} (a) with a different strength $g$. $I_{j}$
with the different $g$ is shown in Fig. \ref%
{wave_pocket_transport_repulsion} (b). It indicates that the interactions
have a significant effect on the particle transport. $M_{1}$ as a function
of $g$ is shown in Fig. \ref{wave_pocket_transport_repulsion} (c). Two
different effects affect the particle transport for the repulsive
interactions. One is that the repulsive interactions always favor the particle
spreading. The other is that the repulsive interactions modify the lattice
intensity equivalently, which is detrimental to particle transport. So the
particle transport $M_{1} $ is a non-monotonic function of $g$.

We have phenomenally proposed the two-body attractive and three repulsive
interactions to understand the exciton patterns. To study how the complex
interactions modify the exciton transport, we show the result with the effective interaction
potential $-g_{1}|\psi _{j}|^{2}+g_{2}|\psi _{j}|^{4}$ in Fig. \ref%
{wave_pocket_transport_repulsion} (d). Here $g_{2}=\epsilon g_{1}^{2}$ with $%
\epsilon =8$. For the large $g_{1}$ cases, the excitons are in repulsive
interactions as that in Fig. \ref{wave_pocket_transport_repulsion} (a).
So $I_{j}$ in Fig. \ref{wave_pocket_transport_repulsion} (e) and $M_{1}$ in
Fig. \ref{wave_pocket_transport_repulsion} (f) are similar to those of
Fig. \ref{wave_pocket_transport_repulsion} (b) and $M_{1}$ in Fig. \ref%
{wave_pocket_transport_repulsion} (c) respectively.

\begin{figure}[htbp]
\includegraphics[width=8cm]{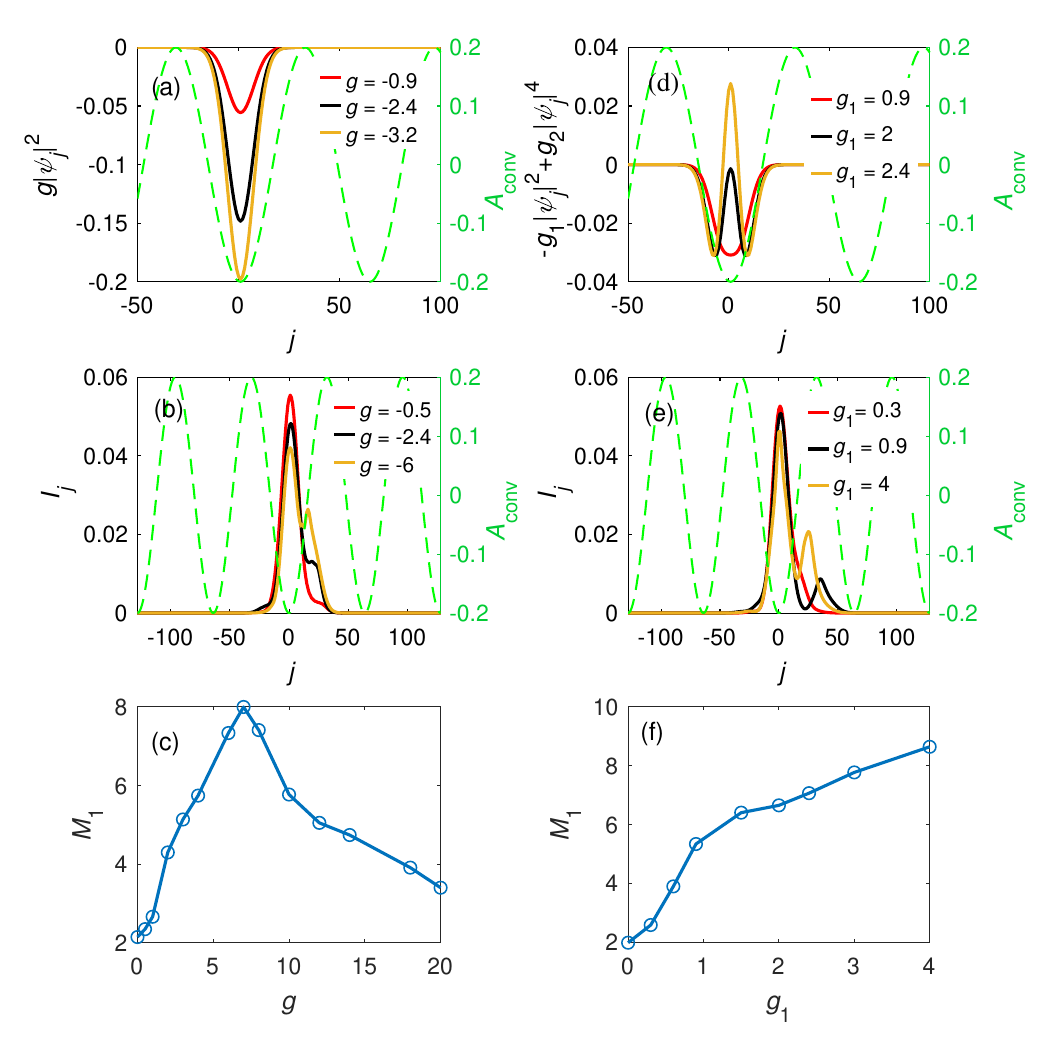}
\caption{The wave pocket diffusions of the attractive particles. The
following parameters are used in the numerical analysis: $A_{\text{conv}} =
0.2$, $v_{\text{conv}} = 2.08$, $\protect\tau = 0.5$ and $\protect\gamma =
0.05$. For the pure attractive case, the interaction parameter $g$%
-dependent effective potential $-g|\protect\psi_j|^2$, the
distribution patterns $I_j$ with different $g$ and transport distance $M_1$
as a function of $g$ are given in (a), (b) and (c) respectively. For the
complex interaction $-g_1 |\protect\psi_j|^2 + g_2 |\protect\psi_j|^4 $ in
the weak attractive interaction region, we set $g_2 = \protect\epsilon %
g_1^2 $ and $\protect\epsilon = 8$. The interaction parameter $g_1$%
-dependent effective potential, the distribution patterns $I_j$ with
the different $g_1$ and transport distance $M_1$ as a function of $g_1$ are
given in (d), (e) and (f) respectively. }
\label{wave_pocket_transport_attraction}
\end{figure}

In the low exciton density case, $-g_{1}|\psi _{j}|^{2}+g_{2}|\psi _{j}|^{4}$
is in the attractive interaction region. It is interesting to study how the
attractive interactions $-g|\psi _{j}|^{2}$ and $-g_{1}|\psi
_{j}|^{2}+g_{2}|\psi _{j}|^{4}$ modify the particle transport. We present
the effective attractive interactions in Figs. \ref%
{wave_pocket_transport_attraction} (a) and (d). In order to study the effects of the
weak interaction more clearly, a small lattice intensity $A_{\text{conv}%
}=0.2$ is adopted. This is in contrast to the repulsive case shown in Figs. \ref%
{wave_pocket_transport_repulsion} (a) and (d) [$A_{\text{conv}}=0.8$]. In
the pure attractive case $-g|\psi _{j}|^{2}$, a wave pocket will collapse
in the time evolution as is shown in Fig. \ref%
{wave_pocket_transport_attraction} (b). However, the finite exciton lifetime
prevents the pocket from collapsing further.

The attractive interactions also have two effects on particle
transport. One is that the attractive interactions always hinder the
particle spread. The other is that the increase of the effective lattice
intensity facilitates the particle transport. So the particle transport $M_1$
is also a non-monotonic function of $g$ [see Fig. \ref%
{wave_pocket_transport_attraction} (c)]. In the small $g_1<4$ case, the weak
complex interactions are in the attractive region, which also
modify the exciton transport in Fig. \ref{wave_pocket_transport_attraction}
(e) and (f). However, $M_1$ shows a monotonic variation as a function of $g_1$.

\subsection{The disorder effects}

\label{The disorder effects}

The data of the exciton transport distance $M_1$ via conveyer as a function
of the conveyer amplitude $A_\text{conv}$ are given in Fig.\ref%
{PRL_106_196806_2011} (g). It shows that the exciton cloud extension $M_1$
is not affected by the motion of a shallow conveyer. Across the transition point however,
the exciton cloud starts to follow the
conveyer, and $M_1$ increases with the increase of $A_\text{conv}$. At higher conveyer amplitude, $M_1$ tends to saturation. As we know, the random disorders or impurities exist inevitably in the
coupled quantum well grown with molecular beam epitaxy. The
destructive interference of scattered waves due to the strong disorders
leads to the Anderson localizations \cite{PhysRev.109.1492, Abrahams}. As a
result, the experimental data is naturally explained as the dynamical
localization-delocalization transition. The dynamical localization was found
in two-band system driven by the DC-AC electric field, in which the Rabi
oscillation is quenched under the certain ratio of Bloch frequency and AC
frequency \cite{PhysRevB.54.R5235, LIU2003301}. An interesting issue here is
whether the data can really be explained by the dynamical localization or not.

The Anderson localizations are generally studied with indirect method where
the random disorders are simulated by a quasiperiodic on-site modulation
\cite{PhysRevLett.110.176403, SUN2023129043}. The quasiperiodic potential is
set to be site dependent, i.e.
\begin{equation*}
V_{\text{dis},j}=2A_{\text{dis}}\cos (2\pi \alpha j),
\end{equation*}%
with $A_{\text{dis}}$ being the strength, $\alpha $ being an irrational
number which is used to characterize the quasiperiodicity. The irrational
number is usually taken as the value of the inverse of golden ratio [$\alpha
=(\sqrt{5}-1)/2$]. The value is closely related to the
Fibonacci number which is defined by $F_{n}/F_{n+1}$ where the Fibonacci
sequence of numbers $F_{n}$ is defined using the recursive relation with the
seed values $F_{0}=0$ , $F_{1}=1$ and $F_{n}=F_{n-1}+F_{n-2}$ \cite%
{PhysRevLett.51.1198, SUN2023129043}. The merit of using Fibonacci numbers
is that the quasiperiodic potential $V_{\text{dis}}\left( j+F_{n+1}\right)
=V_{\text{dis}}\left( j\right) $ becomes periodic.
%In numerical simulations, a
%finite (yet arbitrarily large) number of lattice sites $L=F_{n+1}$ is
%usually assumed to eliminate the domain wall of the PBC and to reduce the
%finite-size effects.XXXXCC

To simplify the discussion, the particle interaction, lifetime and cooling
are not taken into account. The Hamilton in Eq. (\ref{The tight-binding
Hamiltonian}) can be written as
\begin{equation}
H\left( t\right) =\sum_{j}\left[ \tilde{t}\left( \left\vert \psi
_{j}\right\rangle \left\langle \psi _{j+1}\right\vert +\text{h.c.}\right) +%
\mathcal{V}_{j}\left( t\right) \left\vert \psi _{j}\right\rangle
\left\langle \psi _{j}\right\vert \right] ,
\label{The AAH model modulated by the moving lattices}
\end{equation}%
here $\mathcal{V}_{j}\left( t\right) =V_{\text{conv},j}(t)+V_{\text{dis},j}$%
, which is a spatial periodic $\lambda F_{n+1}$ and time periodic $T$
function $\mathcal{V}_{j}\left( t\right) =\mathcal{V}_{j}\left( t+T\right) =%
\mathcal{V}_{j+\lambda F_{n+1}}\left( t\right) $ here $\lambda$ is an integer. The time dependence of the
Hamiltonian (\ref{The AAH model modulated by the moving lattices})
leads to no stationary states in the system. However, the Hamiltonian
(\ref{The AAH model modulated by the moving lattices}) has the time
periodicity. We can write the state as%
\begin{equation*}
\left\vert \psi _{j}\left( t\right) \right\rangle =e^{-i\varepsilon
t}\left\vert c_{j}\left( t\right) \right\rangle,
\end{equation*}%
where $\varepsilon $ is the quasienergy and $\left\vert c_{j}\left( t\right)
\right\rangle =\left\vert c_{j}\left( t+T\right) \right\rangle $ is periodic
function which satisfies the time-dependent equation
\begin{equation*}
\left[ H\left( t\right) -i\partial _{t}\right] \left\vert c_{j}\left(
t\right) \right\rangle =\varepsilon \left\vert c_{j}\left( t\right)
\right\rangle
\end{equation*}%
according to the Floquet theorem \cite{GRIFONI1998229, LIU2003301}. As $%
e^{i\varepsilon \left( t+T\right) }=e^{i\varepsilon t}$, it requires the
quasienergy satisfies $\varepsilon _{l}=\frac{2\pi l}{T}=\omega l$ where $l$ is an
integer. To ensure the consistency between the Floquet theorem and the Bloch
theorem in the form, the period is assumed to be $T=Na$ and the quasienergy
is confined in the first time Brillouin zone $\left] -{\pi }/{a},{\pi }/{a}%
\right] $, with $a$ being the time unit. $\left] -{\pi }/{a},{\pi }/{a}%
\right] $ means an excluded value $-{\pi }/{a}$ on the left and included ${%
\pi }/{a}$ on the right. As a result, $l$ can be taken as $\left] -{N}/{2},{N%
}/{2}\right] $. From the time-dependent nonlinear Schr\"{o}dinger equation,
the Hamiltonian in Eq. (\ref{The AAH model modulated by the moving lattices}%
) can be transformed to a tight-binding Floquet operator in the second
quantization form 
\begin{eqnarray}
\mathcal{H}\left( t\right) &=&\sum_{j}\left[ \tilde{t}\left( c_{j}^{\dag
}\left( t\right) c_{j+1}\left( t\right) +\text{h.c.}\right) \right.
\label{Floquet operator} \\
&&+\left. c_{j}^{\dag }\left( t\right) \left( \mathcal{V}_{j}\left( t\right)
-i\partial _{t}\right) c_{j}\left( t\right) \right] .  \notag
\end{eqnarray}

After the Fourier transform
\begin{eqnarray*}
c_{j}\left( t\right) &=&N^{-1/2}\sum_{n=-N/2}^{N/2}c{}_{j,n}e^{i\omega nt},
\\
c_{j}^{\dagger }\left( t\right)
&=&N^{-1/2}\sum_{n=-N/2}^{N/2}c{}_{j,n}^{\dagger }e^{-i\omega nt},
\end{eqnarray*}%
and the inner product $\left\langle \left\langle \mathcal{H}\left( t\right)
\right\rangle \right\rangle =\frac{1}{T}\int_{0}^{T}\mathcal{H}\left(
t\right) dt=\frac{1}{N}\sum_{n=-N/2}^{N/2}\mathcal{H}\left( n\right) $, the
time-dependent Floquet operator $\mathcal{H}\left( t\right) $ becomes \cite%
{GRIFONI1998229, LIU2021114871}
\begin{eqnarray}
\mathcal{H} &\mathcal{=}&\sum_{j,n}\left\{ \left( \tilde{t}%
c{}_{j,n}^{\dagger }c{}_{j+1,n}+A_{\text{conv}}e^{-i2\pi \eta
j}c{}_{j,n-1}^{\dagger }c{}_{j,n}+\text{h.c.}\right) \right.  \notag \\
&&+\left. \left[ A_{\text{dis}}\cos \left( 2\pi \alpha j\right) -n\omega %
\right] c{}_{j,n}^{\dagger }c{}_{j,n}\right\}.
\label{The time-mean tight binding Hamiltonian}
\end{eqnarray}
It is obviously that several sub-bands forming for different $n$ of the model
(\ref{The time-mean tight binding Hamiltonian}). For a given $n$,
the model (\ref{The time-mean tight binding Hamiltonian}) can be
mapped to Harper model which is the two-dimensional electron gas in a
magnetic field if $\eta$ is regarded as a magnetic flux penetrating the unit
cell \cite{Harper_1955}.

Under the Floquet ansatz $\left\langle \left\langle c_{j}^{\dag
}\left( t\right) c_{j}\left( t\right) \right\rangle \right\rangle
=\sum_{n}c{}_{j,n}^{\dagger }c{}_{j,n}$, we can define a time-average
inverse of the participation ratio (TMIRP) of the normalized eigenstate $%
\left\vert c{}_{j,n}\right\rangle _{l}$ corresponding to the eigenvalue $%
\varepsilon_{l} $,
\begin{equation*}
\text{TMIPR}_l=\sum\limits_{j,n}\left\vert _{l}\left \langle
c{}_{j,n}^{\dagger }|c{}_{j,n}\right\rangle _{l}\right\vert ^{4}.
\end{equation*}%
The localization of the whole system can be characterized by the average of $%
{\text{TMIPR}}$
\begin{equation*}
\overline{\text{TMIPR}}=\sum_{l}{\text{TMIPR}_{l}}/L.
\end{equation*}
For a delocalizaton phase of the system, $\overline{\text{TMIPR}} $ is of
the order $1/L$, whereas it approaches $1$ in the localized phase.

\begin{figure}[tbph]
\includegraphics[width=7cm]{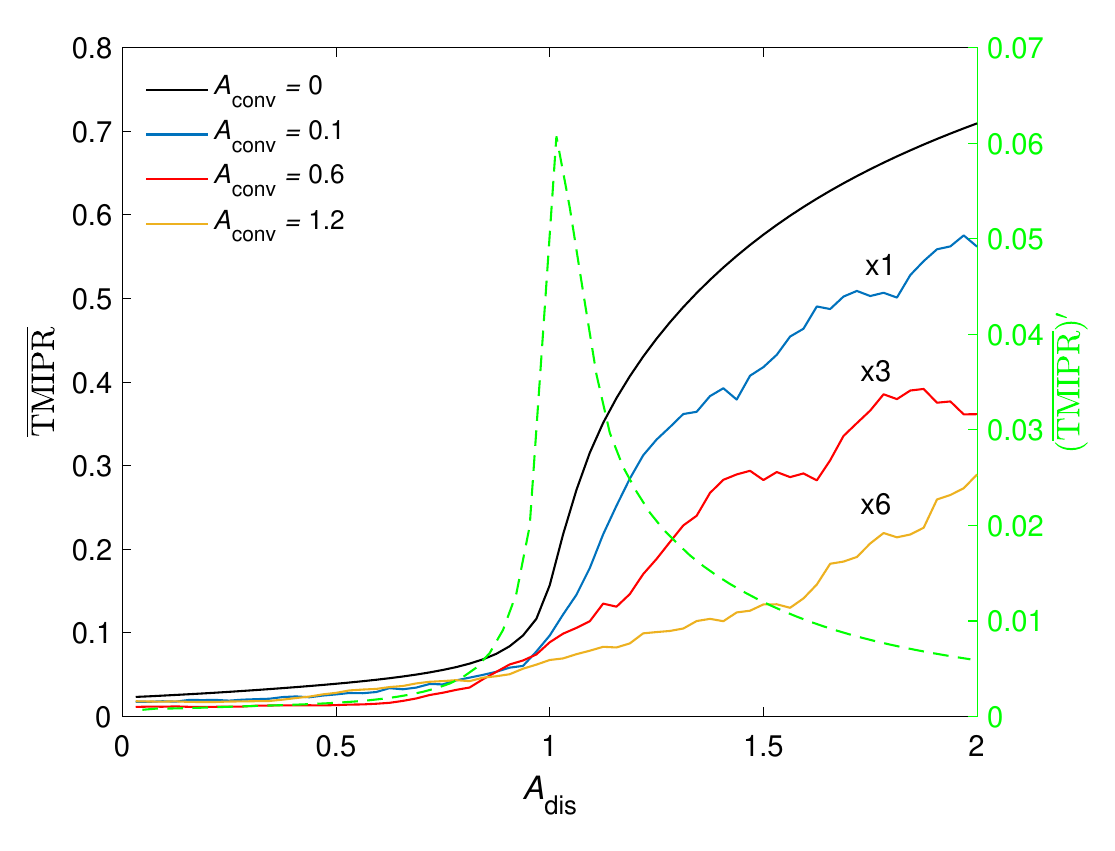}
\caption{The $\overline{\text{TMIPR}}^{\prime }$s dependence on $A_{\text{dis%
}}$ for different $A_{\text{conv}}$. The data have been expanded three and
six times for the cases of $A_{\text{conv}}=0.6$ and $1.2$. The green dashed
line is the slop of black line and is used to show the phase transition
occurring at $A_{\text{dis}}=1$ for the AAH model ($A_{\text{conv}}=0$). The
parameters used in the numerical calculation are $L=N=65$, $\protect\alpha %
=0.2$ and $\protect\omega =0.5$.}
\label{TMIPR}
\end{figure}

We diagonalize the time-mean tight-binding Hamiltonian (\ref{The
time-mean tight binding Hamiltonian}) to get the eigenstates $\left\vert
c{}_{j,n}\right\rangle _{l}$ for calculating $\overline{\text{TMIPR}}$. The result is shown
in Fig. \ref{TMIPR}. In the case of $A_{\text{conv}}=0$, the Hamiltonian (\ref{The AAH model modulated by the moving lattices}) is reduced to the
Aubry-Andr\'{e} -Harper (AAH) model \cite{AIPS.3.133, Harper_1955}. The $A_{%
\text{dis}}$ dependence of $\overline{\text{TMIPR}}$ and its first derivative $\left(
\overline{\text{TMIPR}}\right) ^{\prime }=d\overline{\left( \text{TMIPR}%
\right) }/d\left( A_{\text{dis}}\right) $ are shown in Fig. \ref{TMIPR} by
the black line and the green dashed line, respectively. A peak is found at $A_{\text{dis}%
}=1$ which is consistent with that of the AAH model where the Anderson
localization transition occurs at $A_{\text{dis}}= \tilde{t}$. It indicates the
effectiveness of the time-mean method. We further calculate the $A_{\text{dis%
}}$ dependence of $\overline{\text{TMIPR}}$ for different $A_{\text{conv}}$.
Similar to the above research, for a wave pocket driven by the moving lattices, the wave
pocket spreads over time, and tends to delocalization. For a wave pocket driven by the random lattice, it tends to localize. As a result, the interplay of
the two lattices leads to suppression of the $\overline{\text{TMIPR}}$ with
the increase of $A_{\text{conv}}$. As we know, when there is a competition between
localization and delocalization in particular, a localization\mbox{-}%
delocalization transition usually occurs. It is surprising that no visible
localization\mbox{-}delocalization transition is found in the $A_{\text{dis}}
$ dependence of $\overline{\text{TMIPR}}$ in Eq. (\ref{TMIPR}) for the case
of $A_{\text{conv}} \geq 0.6$.

The localization\mbox{-}delocalization transition 
%of the AAH model [$A_{%
%\text{conv}}=0$ in Eq. (\ref{The AAH model modulated by the moving lattices}%
%)] 
can also be captured by the asymptotic behavior over long time of the
second-order moment of position operator $\sigma ^{2}(t)$ defined by the
wave spreading \cite{PhysRevB.50.1420, JournalPhysA.31.1353,
PhysRevB.103.054203}
\begin{equation*}
\sigma ^{2}(t)=\sum_{j}\left\langle c_{j}(t)\right\vert j^{2}\left\vert
c_{j}(t)\right\rangle,
\end{equation*}%
with the initial localized state $\left\vert c_{j}(0)\right\rangle $$=\delta
_{j.0}$. 
It gives the opportunity to further check the non-existence of the localization\mbox{-}delocalization transition.
The asymptotic spreading of $\sigma ^{2}(t)$ is described by the
power law, i.e. $\sigma ^{2}(t)\sim t^{2\delta }$ where $\delta $ is the
diffusion exponent. In terms of dynamical behavior of a wave pocket,
measured by the exponent $\delta =\delta (A_{\text{dis}})$, the phase
transition is discontinuous since $\delta (A_{\text{dis}})=1$ for $A_{\text{%
dis}}$ $<\tilde{t}$ (ballistic transport), $\delta (A_{\text{dis}})\simeq
1/2 $ at the critical point $A_{\text{dis}}=\tilde{t}$ (almost diffusive
transport), and $\delta (A_{\text{dis}})=0$ in the localized phase $A_{\text{%
dis}}>\tilde{t}$ (dynamical localization). When the spreading velocity
$v(A_{\text{dis}})\sim \sigma (t)/t$ is further defined, the phase
transition turns out to be smooth, with $v(A_{\text{dis}})$ being continuous
function of potential amplitude $A_{\text{dis}}$ and $v(A_{\text{dis}})=0$
for $A_{\text{dis}}\geq 0.6$.

\begin{figure}[htbp]
\includegraphics[width=9cm]{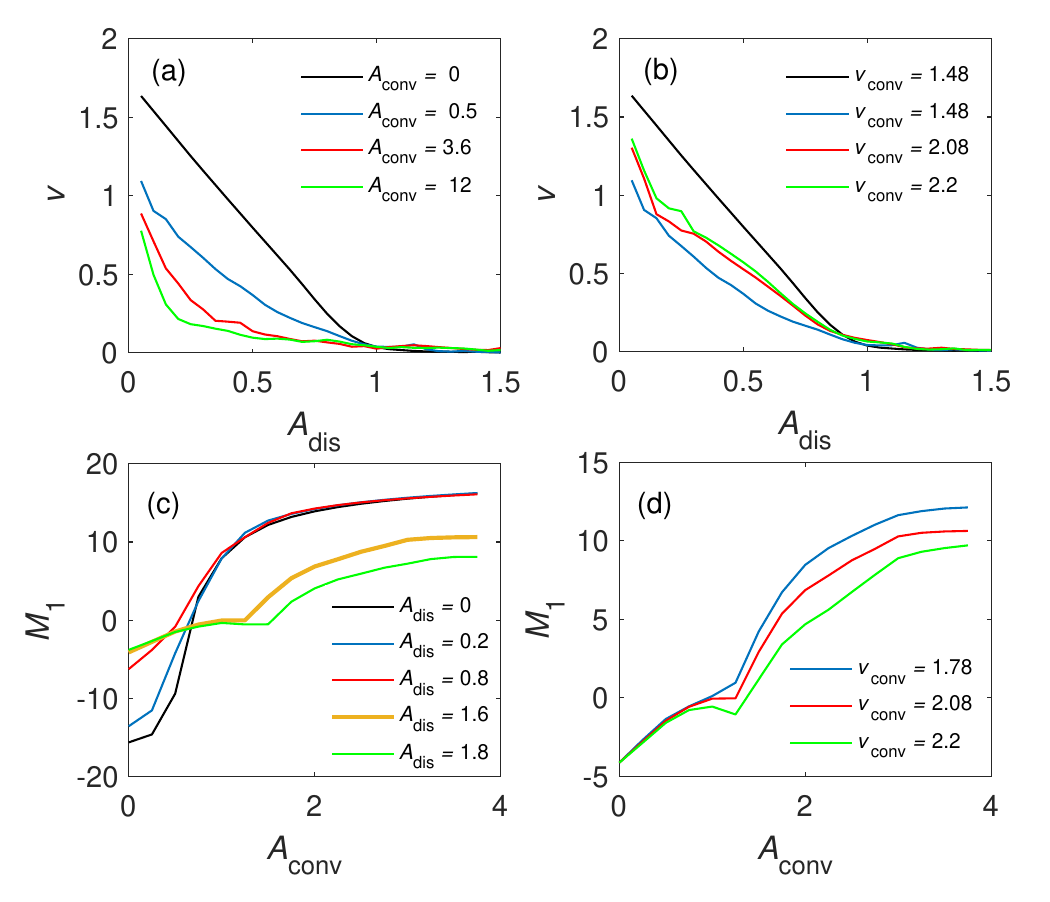}
\caption{$v$ as a function of the disorder amplitude $A_{\text{dis}}$ for
the largest propagation time $t = 150$ where in (a) the speed of the moving
lattice is set to be $v_{\text{conv}}=1.48$ and in (b) the disorder amplitude
is set to be $A_{\text{conv}}=1.6$ respectively. ${M_1}^{\prime }$s
dependence on the conveyer amplitude $A_{\text{conv}}$ for the different
disorder intensity $A_{\text{conv}}$ [$v_{\text{conv}}=2.02$] (c) and for
the different conveyer speed $v_{\text{conv}}$ [$A_{\text{conv}}=1.6$] (d).
The other parameters $\protect\gamma=0.05$ and $\protect\tau = 0.5$.}
\label{localization}
\end{figure}

We still use the above method to characterize the dynamical localization. We
solve the time-dependent nonlinear Schr\"{o}dinger equation of
the Hamiltonian (\ref{The AAH model modulated by the moving lattices}%
) with the initial localized state $\left\vert c_{j}(0)\right\rangle $ $%
=\delta _{j.0}$ to obtain $\left\vert c_{j}(t)\right\rangle $ and $v(A_{%
\text{dis}})$. The dependence of $v(A_{\text{dis}})$ on $A_{\text{dis}}$
is shown in Fig. \ref{localization}. For the case of $A_{\text{dis}}=0$,
the Hamiltonian in Eq. (\ref{The AAH model modulated by the moving lattices}%
) is reduced to the AAH model. The dependence of ${A_{\text{conv}}}$ on $%
v(A_{\text{dis}})$ [black line] shows a transition at $A_{\text{dis}}=1$$(%
\tilde{t})$ obviously. With the applying of the moving lattices, either increasing its
intensity $A_{\text{conv}}$ [$v_{\text{conv}}$ unchanged] in Fig. \ref%
{localization} (a) or increasing its velocity $v_{\text{conv}}$ [$A_{\text{%
conv}}$ unchanged] in Fig. \ref{localization} (b), the spreading velocity $%
v(A_{\text{dis}})$ is suppressed. The most obvious feature is the
disappearance of the delocalization\mbox {-}localization transition. We also
calculate the conveyer amplitude ${A_{\text{conv}}}^{\prime }$s dependence
on $M_1$ for the different disorder intensity $A_{\text{dis}}$ [$v_{\text{%
conv}}=2.02$] in Fig.\ref{localization} (c) and for the different conveyer
velocity $v_{\text{conv}}$ [$A_{\text{dis}}=1.6$] (d). Recalling the
behavior of $\overline{\text{TMIPR}}$ in Fig. \ref{TMIPR}, we therefore argue
that the moving lattice breaks the Anderson localization transition.

\subsection{Comparing the theory with the experiments}

\begin{figure}[htbp]
\includegraphics[width=8cm]{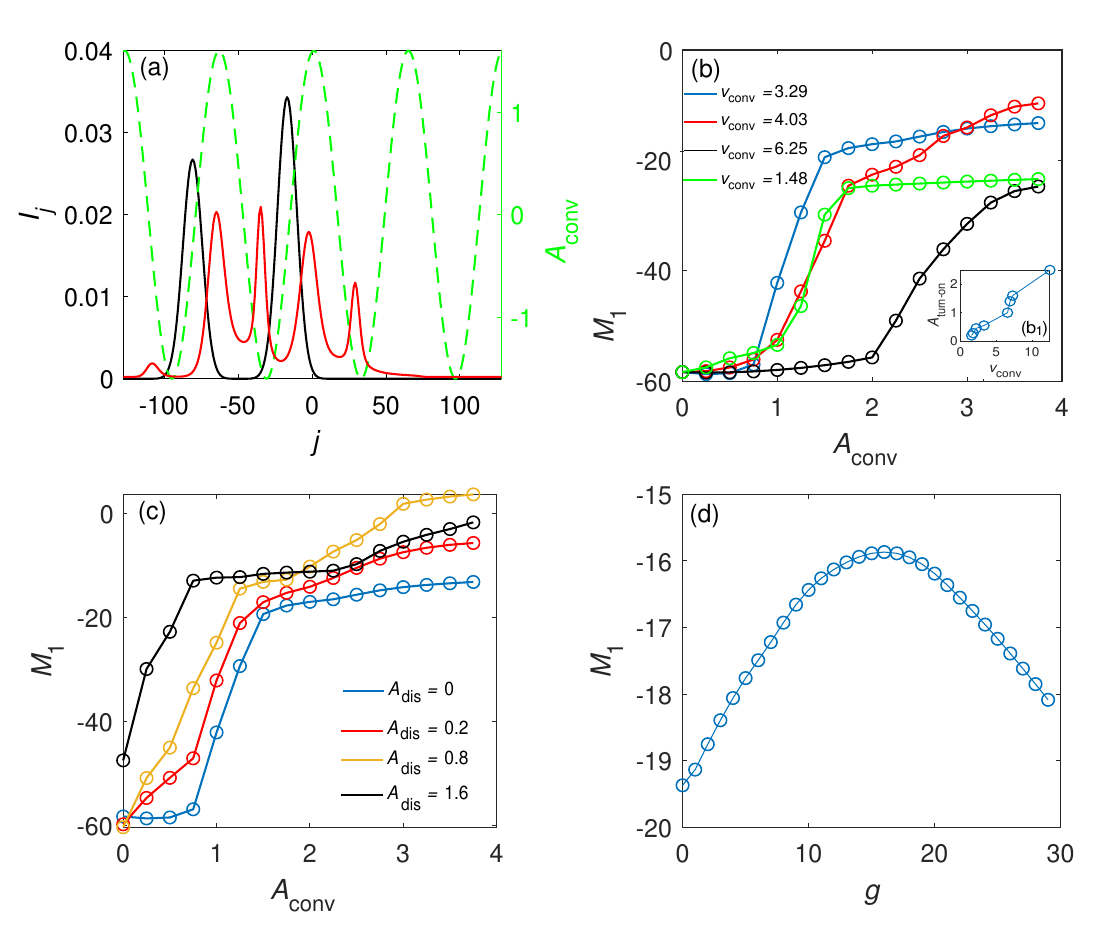}
\caption{(a) Red line: the PL intensity profile $I_j$ . Black line: the
initial wave-pocket. Green-dashed line: the initial conveyer. The parameters
$v_{\text{conv}} =1.48$ and $A_{\text{conv}} = 1.6$ are used in the
calculations. (b) $M_{1}$ as a function of the conveyer amplitude $A_{\text{%
conv}}$ for the different conveyer velocity $v_{\text{conv}} =$ 1.48, 3.29,
4.03 and 6.25 respectively. (b$_{1}$) $A_{\text{turn-on}}$ versus the
conveyer velocity $v_{\text{conv}}$. (c) $M_{1}$ as a function of the
interaction parameter $A_{\text{conv}}$. (d) $M_{1}$ as a function of the
interaction parameter $g$ with the parameters $v_{\text{conv}} = 4.03$, $A_{%
\text{conv}} = 2.5$. The other parameters in these calculations are set as $g = 1.2$,$%
\protect\tau =0.5 $ and $\protect\gamma = 0.05$.}
\label{Comparing_with_the_experiments}
\end{figure}

After detailed investigations of the various effects on the diffusion of a wave pocket in the
moving lattice, we are ready to study the experimental data. According to
the discussions in sec. \ref{The disorder effects}, the plateaus can not
be explained by delocalization\mbox
{-}localization transition due to the disorders, even though the disorders
suppress the wave-pocket transport. The disorders are firstly neglected in
following numerical calculations. In addition, as the coherent excitons have
different origins, the wave function of the coherent excitons was assumed
with two peaks initially. The motivation of the two peaks comes from the two
rings reported in Refs. \cite{Butov2002a, Butov2002b}. We solve the
time-dependent nonlinear Schr\"{o}dinger equation (\ref{time-dependent
nonlinear Schrodinger equation}) with the initial wave function [black line
in Fig. \ref{Comparing_with_the_experiments}] to obtain $I_{j}$ [red line in
Fig. \ref{Comparing_with_the_experiments} (a)]. The appearance of the four
peaks is basically consistent with the interference fringe.

The conveyer amplitude ${A_{\text{conv}}}^{\prime }$s dependence on $M_{1}$
is shown in Fig. \ref{Comparing_with_the_experiments} (b) for different $v_{%
\text{conv}}$. It indicates that the transport distance $M_{1}$ increases
with the conveyer amplitude $A_{\text{conv}}$. However, the exciton
transport is less efficient for higher velocity. We also study the
low-velocity case [green line] and find that the exciton transport is also
less efficient for lower velocity. It indicates that the effective transport
occurs at moderate conveyer speed.

Although no disorders are involved in the potential, the plateau can still
be found in the low moving lattice amplitude $A_{\text{conv}}$. In
particular, the plateau width increases with the conveyer velocity $v_{%
\text{conv}}$ [see Fig. \ref{Comparing_with_the_experiments} (b$_1$)].
This is consistent with the experimental data [given by black points in
 Fig. \ref{PRL_106_196806_2011} (g)] where $A_{\text{turn-on}}$
[defined as $A_{\text{conv}}$ at the line intersection] increases with $v_{%
\text{conv}}$. The result with the disorders is shown in Fig. \ref%
{Comparing_with_the_experiments} (c), it is interesting to see that the
plateau is destroyed with the increase the disorder intensity $A_{\text{dis}
} $. We therefore believe that the plateau cannot be caused by the
dynamical localization-delocalization transitions due to the disorders. We
infer that the experimental sample is very clean. Their random potential
strength is very weak, even if impurities exist.

According to the discussion in sec. \ref{A wave pocket transport with the
particle interactions}, there is no difference between $g|\psi_j|^2$ and $%
-g_1|\psi_j|^2 + g_2|\psi_j|^4$ in describing the exciton interactions except
for the case of the very low particle density. 
We use the interaction $g|\psi_j|^2$ to calculate the $g$
dependence of $M_{1}$ as is shown in Fig. \ref{Comparing_with_the_experiments}
(d). The transport distance $M_{1}$ increases firstly and then decreases
with $g$. The behavior is similar to the experimental data in Fig. \ref%
{PRL_106_196806_2011} (h) where the exciton transport via conveyer $M_1$
increases firstly and then decreases with the excitation power $\log (P_{%
\text{ex}})$. As $g\propto N$ when the wave function $\psi_j$ is normalized,
it indicates that the coherent exciton number $N\propto \log (P_{\text{ex}})$%
. We therefore infer that it is less efficient to increase the coherent
exciton number by increasing the laser power $P_{\text{ex}}$. The cooling
speed and long lifetime are still the key factors.

\section{Summary}

\label{Summary}

In summary, we have investigated the experimental data of electrostatic
conveyor belt for excitons to understand the dynamical behaviors. We found
that the formation of exciton patterns come from both the spatially-separated hot excitons and cooled excitons. The hot excitons can be regarded as classical
particles whose transport can be well described by the classical diffusion
equation. However, the cooled excitons are the coherent Bosons which must be
described by the Schr\"{o}dinger equation. The studies capture the nature
of the exciton diffusion in conveyor belt, i. e. the excitons are cooled
down during the transport. In particular, the excitons are in highly
degenerate states far from the laser spot, and their temperature becomes
lower and lower. This is why the method of the real-time and imaginary-time
evolution of the Schr\"{o}dinger equation can give a good account of the
spatial separation patterns. Even though some discrepancies exist between
the theory and experiment, the numerical results are expected to be realized
in the further.

By calculating the distribution of the escape probability, we get the PL
stripes and the transport distance that are consistent with the experiment. We
find the cooling speed is the key factor to the transport distance.
Based on the comparison between the calculation and the experimental data of
transport distance, a function of density, we find increasing the
excitation power is not an effective approach to increase the coherent
excitons. We also find that the disorders failed to induce the dynamical
localization-delocalization transition in the moving lattices. As a result,
we infer the sample is basically impurity free.

In order to realize the controllable exciton transport in the moving
lattices, according to our studies, two priority research directions are
obvious. Experimental study should still focus on finding the ways of the
long lifetime and the fast cooling speed of the excitons. Theoretically,
designing achievable moving lattice to realize the dynamical
localization-delocalization transition may be a new research direction.
After finishing the manuscript, we are aware of the experiment of transport
and localization of indirect excitons in a van der Waals MoSe$%
_2$/WSe$_2$ heterostructure \cite{fowlergerace2023transport}. Whether the
theory present in the manuscript can explain the data or not is also worthy of further theoretical study.

\begin{acknowledgments}
This work was supported by Hebei Provincial Natural Science Foundation of China (Grant
No. A2010001116, A2012203174, and D2010001150),
and National Natural Science
Foundation of China (Grant Nos. 10974169, 11174115 and 10934008).
\end{acknowledgments}

\bibliography{Refs}

\end{document}